\documentstyle[preprint,aps,prb,psfig]{revtex}

\include{epsfig}

\tighten

\begin{document}

\draft

\preprint{HD-11}

\title{Kinetic Ising System in an Oscillating External Field: Stochastic
Resonance and Residence-Time Distributions}

\author{S.W. Sides,$^{\ast \dagger}$
R.A. Ramos,$^{\ast \dagger \circ}$ P.A. Rikvold,$^{\ast \dagger}$
and M.A. Novotny $^{\dagger \ddagger}$}

\address{$^{\ast}$Center for Materials Research and Technology and Department of Physics, and\\
$^{\dagger}$Supercomputer Computations Research Institute, \\
Florida State University, Tallahassee, Florida 32306-3016 \\
$^{\circ}$Department of Physics, University of Puerto Rico, Mayaguez, PR 00681 \\
$^{\ddagger}$Department of Electrical Engineering \\
FAMU--FSU College of Engineering, Tallahassee, Florida 32310-6046}

\maketitle

\begin{abstract}

Experimental, analytical, and numerical results
suggest that the mechanism
by which a uniaxial single-domain ferromagnet
switches after
sudden field reversal depends
on the field magnitude and
the system size.
Here we report new results on how these
distinct
decay mechanisms influence hysteresis in a
two-dimensional
nearest-neighbor kinetic Ising model.
We present theoretical predictions
supported by numerical simulations
for the frequency dependence of the
probability distributions
for the hysteresis-loop area and the
period-averaged magnetization,
and for the residence-time distributions.
The latter suggest evidence of stochastic
resonance for small systems
in moderately weak oscillating fields.
\end{abstract}


\narrowtext

\bibliographystyle{unsrt}

%%\section{Introduction}

The frequency dependence of hysteresis-loop areas,
$A$=$-\oint m dH$, for 
ultrathin ferromagnetic
${\rm Fe/Au(001)}$ films \cite{hysExpt1}
appear consistent with numerical and theoretical
studies of kinetic Ising and mean-field models. \cite{rao,muktishE,tome,lo}
These mean-field models exhibit a dynamic phase transition at
a critical frequency above which the period-averaged magnetization,
$Q$=$(\omega / 2 \pi) \oint m dt$, becomes nonzero. \cite{tome,lo}
It has been suggested that an analogous transition occurs in
spatially extended systems, \cite{rao,muktishE} and the
results of Ref. 1 indicate that it may be experimentally
relevant when spin-waves do not
contribute significantly to the switching dynamics.

Theoretical arguments and numerical simulations indicate the
existence of distinct parameter regimes in which a uniaxial
single-domain ferromagnet, following instantaneous field
reversal, should switch to the stable magnetization direction,
either by nucleation of a {\it single} droplet of the stable
phase [the single-droplet (SD) regime]
or by simultaneous nucleation and growth of {\it many} droplets
[the multi-droplet (MD) regime]. \cite{rtms,howard}
The SD regime corresponds to moderately weak fields and/or
small systems, whereas the MD regime corresponds to moderately
strong fields and/or larger systems.

Here we present new results from our study of the
effects of the SD and MD decay mechanisms on hysteresis. \cite{mmm95}
We are not aware that these effects have been systematically studied
by other groups.
We present data from MC simulations for the $Q$ and $A$
{\it probability distributions} to highlight
novel behavior in the response to a time-dependent field.
We give a theoretical fit for the
frequency dependence of the hysteresis loop
area, which
takes into account knowledge about the dynamics of metastable decay and
is applicable to a wide range of frequencies of the external field.
We also give a theoretical expression for
the residence time distribution (RTD)
that agrees well with the simulation data,
and we show how the
frequency dependence of the RTD
suggests the presence of stochastic resonance.
A full account of our methods and results will be given elsewhere. \cite{sideshys}

The model used in our study is a kinetic, nearest-neighbor
Ising ferromagnet on a
square lattice with periodic boundary conditions.
The Hamiltonian is given by
${\cal H }$=$-J \sum_{ {\em \langle ij \rangle}} {\em s_{i}s_{j}} - H(t) \sum_{i} {\em s_{i}}$,
where $\sum_{ {\em \langle ij \rangle} }$ runs over all
nearest-neighbor pairs, and $\sum_{i}$ runs over all
$N$=$L^{2}$ lattice sites.
The ferromagnetic exchange coupling is $J>0$ and $H(t)$ is a
time-dependent external field.
The dynamic used
is the Glauber single-spin flip Monte
Carlo algorithm,
with updates at randomly chosen sites.
This is
defined by the spin-flip probability \cite{glauber2}
$W(s_{i} \rightarrow -s_{i})$=$\exp(- \beta \Delta E_{i})/(1 + \exp(- \beta \Delta E_{i}))$,
where $\Delta E_{i}$ gives the change in the energy of the system
if the spin flip is accepted,
and $\beta^{-1} = k_{\rm B}T$ is the temperature in energy units.
Time is given in units of Monte Carlo steps per spin (MCSS).
The {\it lifetime} of the metastable phase in a field
of magnitude $|H|$, $\tau(|H|)$, is defined as the average
time it takes the magnetization to reach zero, following
instantaneous field reversal.
The frequency of the sinusoidal applied field,
$H(t) = -H_{o}\sin (\omega t)$,
is chosen by specifying the ratio $R$ of its
period to the metastable lifetime,
$R$=$\frac{(2 \pi / \omega)}{\tau(H=H_{o})}$.

In our simulations, a square system of side $L$=$64$ at $T$=$0.8T_{c}$
is prepared  with $m(0)$=$0$ with the spins in a
random arrangement.
Then the sinusoidal $H(t)$ is applied,
and the magnetization $m(t)$ is recorded for $1.69 \times 10^{7}$ MCSS.
The field amplitude used for the simulations in the MD regime
is $H_{o}$=$0.3J$ ($\tau$=$75$MCSS) and in the SD regime it
is $H_{o}$=$0.1J$ ($\tau$=$2000$MCSS).
For the values of $L$ and $T$
used in these simulations,
the field strength corresponding to the crossover
between these two regimes is approximately $0.11J$.

%%\section{Results}

Figure \ref{qdis} shows
the probability distributions of the period-averaged
magnetization $Q$, for several values
of $R$.
In the MD regime
the shape changes from bimodal to unimodal
between $R$=$3$ and $4$, possibly
indicating a dynamic phase transition. \cite{tome,lo}
The symmetric, bimodal shape indicates
that the sign of $Q$
changes many times
over the course of a simulation run.
Plots of $m(t)$ show
that this switching occurs slowly,
over several periods of $H(t)$. \cite{mmm95}
In the SD regime,
the probability distributions for all values of $R$,
except for the very largest,
are bimodal with sharp
peaks near $\pm 1$.
This indicates that $m(t)$ remains near $\pm 1$
for many consecutive periods, punctuated by
abrupt switching events.
A unimodal distribution in the SD regime
would require even larger values of $R$ than shown here.

We have also calculated the probability distributions for the
hysteresis-loop area $A$ in the MD regime for
$R$=$2$ through $200$, corresponding to frequencies of
$6.7 \times 10^{-3}$
through $6.7 \times 10^{-5}$ ${\rm MCSS}^{-1}$.
The widths of these distributions,
as well as the mean values,
depend on $R$, with narrow distributions for
small and large
$R$ and wider distributions at intermediate $R$.
The mean values and standard deviations are plotted versus
frequency in Fig. \ \ref{aMDstat}.
Note that the maximum in the standard deviation of $A$ occurs in a frequency range
close to where a dynamic transition for $Q$ in the MD regime may
be located.
The solid curve is the result of
a calculation using expressions for the volume
of droplets of the stable phase in a time-dependent field,
based on Avrami's law. \cite{sekimoto}
These equations are numerically integrated
to give $m(t)$,
which is then used to calculate the integral in the definition of $A$.
The fit uses three parameters.
Two of these parameters
are obtained from field-reversal simulations that measure $\tau$.
The third parameter
is estimated from measurements of $\tau$
in a time-dependent field
at the very lowest frequency
shown in Fig. \ref{aMDstat},
$1/R$=$0.005$.
Studies of finite-size effects in the MD regime
are in progress. \cite{sideshys}

An example of
the agreement between the theoretical and simulated RTD's
in the SD regime
is shown in Fig.\ref{rtdSD}.
The residence time with no cutoff is defined as the
interval between consecutive times
when $m(t)$ crosses zero.
We use a cutoff of $0.25$
to exclude events when $m(t)$ crosses and recrosses
zero within
a time interval much shorter than the
period of the field oscillation.
The theoretical curve is obtained
by considering the switching as a Poisson process with a variable
rate, identical to the field-dependent nucleation rate.
All the parameters in the theoretical expression were obtained from
field-reversal experiments.
The peak strength $S_{j}$ is defined as the area under the $j$-th
peak in the RTD.
The frequency dependence of the peak strengths
is shown in Fig. \ref{rtdSDpeak}.
A similar analysis
has been used in a study of
stochastic resonance
in a bistable process with Gaussian noise. \cite{SRbonafide}
In that case, the
frequency at which $S_{1}$ has a maximum corresponds to
the mean escape rate due to the noise alone.
As seen in Fig.\ \ref{rtdSDpeak}, $S_{1}$ does not pass through a maximum
although the other $S_{j}$ do show the maxima characteristic of
stochastic resonance.
In contrast to the system studied in Ref. \ 12,
the Ising model used in our simulations is in a parameter
regime where the thermal relaxation
time at zero applied field is practically infinite.
Reasonable switching times are obtained only when
both the applied field and thermal
fluctuations are present.
For $S_{1}$ to display a maximum, the switching time due
to thermal fluctuations alone would have to be smaller than
the period of $H(t)$.
As a consequence, there is no maximum in $S_{1}$ for
practically attainable frequencies.

In summary, our results show distinct differences
between the MD and SD regimes with respect to the response
of a kinetic Ising system to an oscillating
external field.
We present a good fit to the average hysteresis loop area
in the MD region, using
a droplet picture to describe the decay of the metastable 
phase in a time-dependent field.
This is a one-parameter fit using data in the very low frequency range.
The other parameters
come from simulations and theory for the
instantaneous field-reversal case.
In addition, we show an essentially parameter-free
theoretical result for the RTD's
in the SD regime.
The agreement between theory and simulation is quite good.
The plot of the peak strengths $S_{j}$ vs. frequency shows that
$S_{j}$ for
$j>1$ have maxima
indicative of stochastic resonance. \cite{SRbonafide}
Whereas our data do not have a maximum in $S_{1}$, the maxima
for $j>1$
suggest a competition between a stochastic timescale, the lifetime, and a
deterministic timescale, the period of $H(t)$.

Supported in part by 
FSU-MARTECH,
FSU-SCRI under DOE
Contract No.\ DE-FC05-85ER25000, 
and NSF
Grants No.\ DMR-9315969 and DMR-9520325. 

%\begin{thebibliography}{9}

%%%%%%%%%%%%%%%%%%%%%%%%% FIGURE CAPTIONS %%%%%%%%%%%%%%%%%%%%%%%%%%

%% FIGURE Q DIST MD/SD
\begin{figure}
\caption{\label{qdis}
Probability distributions of the
period averaged
magnetization, $Q$=$(\omega /2 \pi) \oint m dt$,
for (a) the MD regime and (b) the SD regime
for several values of $R$.
Note: the asymmetries in some of the
distributions are due to the
finite length of the simulations.}
\end{figure}

%% FIGURE AREA DIST. MD STAT 
\begin{figure}
\caption{\label{aMDstat}
Mean values of the hysteresis
loop areas $A$, shown vs.\ frequency in the MD regime.
The frequency is given in units of the inverse lifetime
in static field, $1/R$.
The ``error bars'' give the standard deviations for the loop-area distributions.
The solid line is a fit, described in the text.
(Note: the points at $1/R$=$1/12, 1/80, 1/140$ and $1/200$ are calculated from
shorter time series, $6.0 \times 10^{5}$ MCSS.)}
\end{figure}

%% FIGURES SD RES. TIME DIST.
\begin{figure}
\caption{\label{rtdSD}
Residence time distribution for the SD regime with $R$=$10$,
corresponding to a period of $2 \pi / \omega$=$20580$ MCSS.
The time axis is scaled
by $2 \pi / \omega$.
The circles are data from MC simulations.
The solid line is a theoretical result.
Note: The binsize must be multiplied by the period of the field
for the correct normalization.}
\end{figure}

%% FIGURES SD RES. TIME DIST. PEAK STRENGTHS
\begin{figure}
\caption{\label{rtdSDpeak}
Peak strengths, $S_{j}$, vs. frequency for the first four peaks in the
residence time distributions in the SD regime.
The curves are from the same theoretical calculation used for the RTD.
The set of three points near frequency $1/R$=$0.1$
corresponds to the data in Fig. 3.}
\end{figure}
\pagebreak

%%%%% INPUT POSTSCRIPT FIGURES ONE COLUMN FORMAT %%%%%%%%%%%%%%%%%%%%
\begin{figure}
\noindent
Figure 1
\hspace{5.0in}
{\large {\bf HD-11}} \\
S.W. Sides \\
MMM 96 \\
\centerline{\psfig{figure=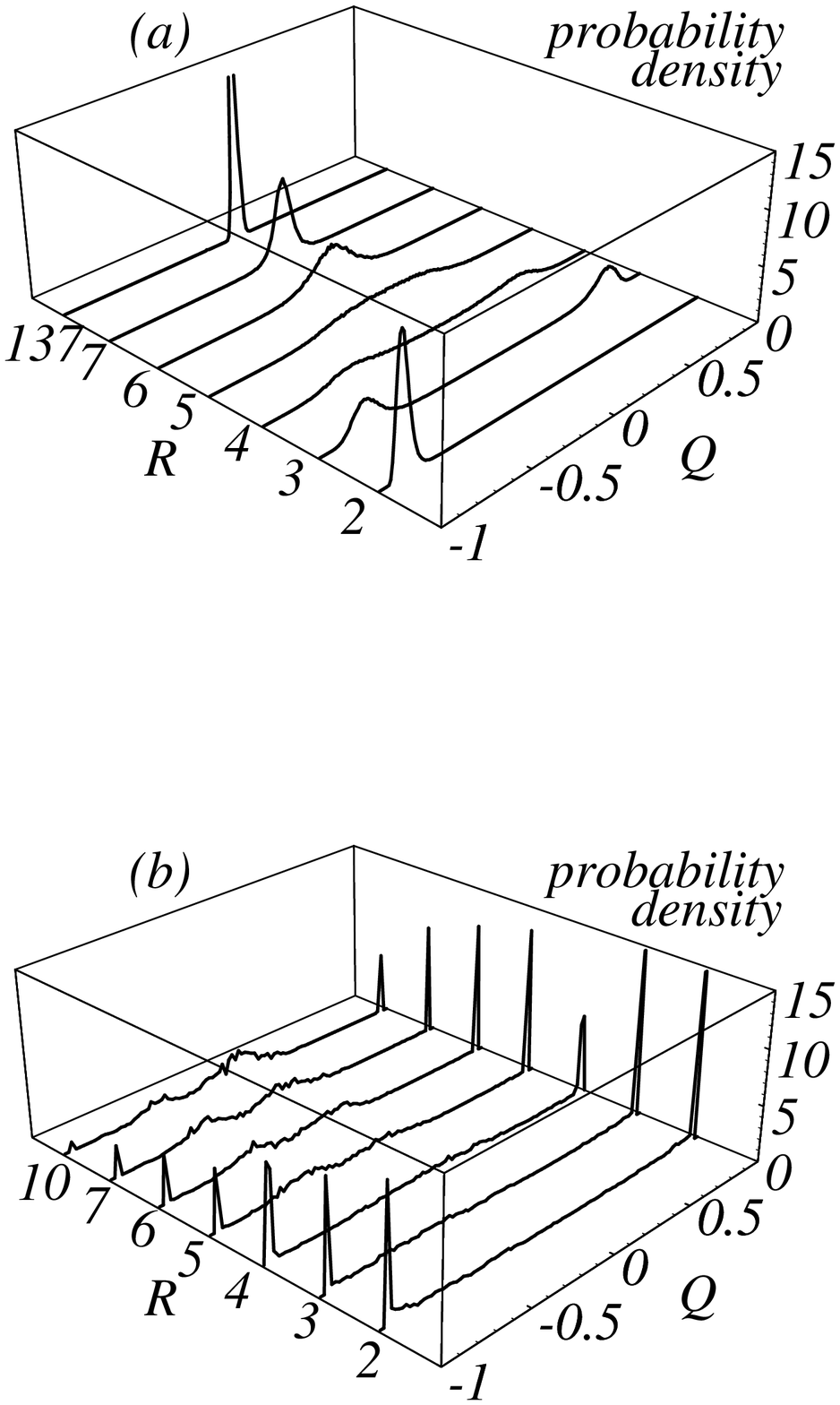,width=5.0in}}
\end{figure}
\pagebreak

\begin{figure}
\noindent
Figure 2
\hspace{5.0in}
{\large {\bf HD-11}} \\
S.W. Sides \\
MMM 96 \\
\centerline{\psfig{figure=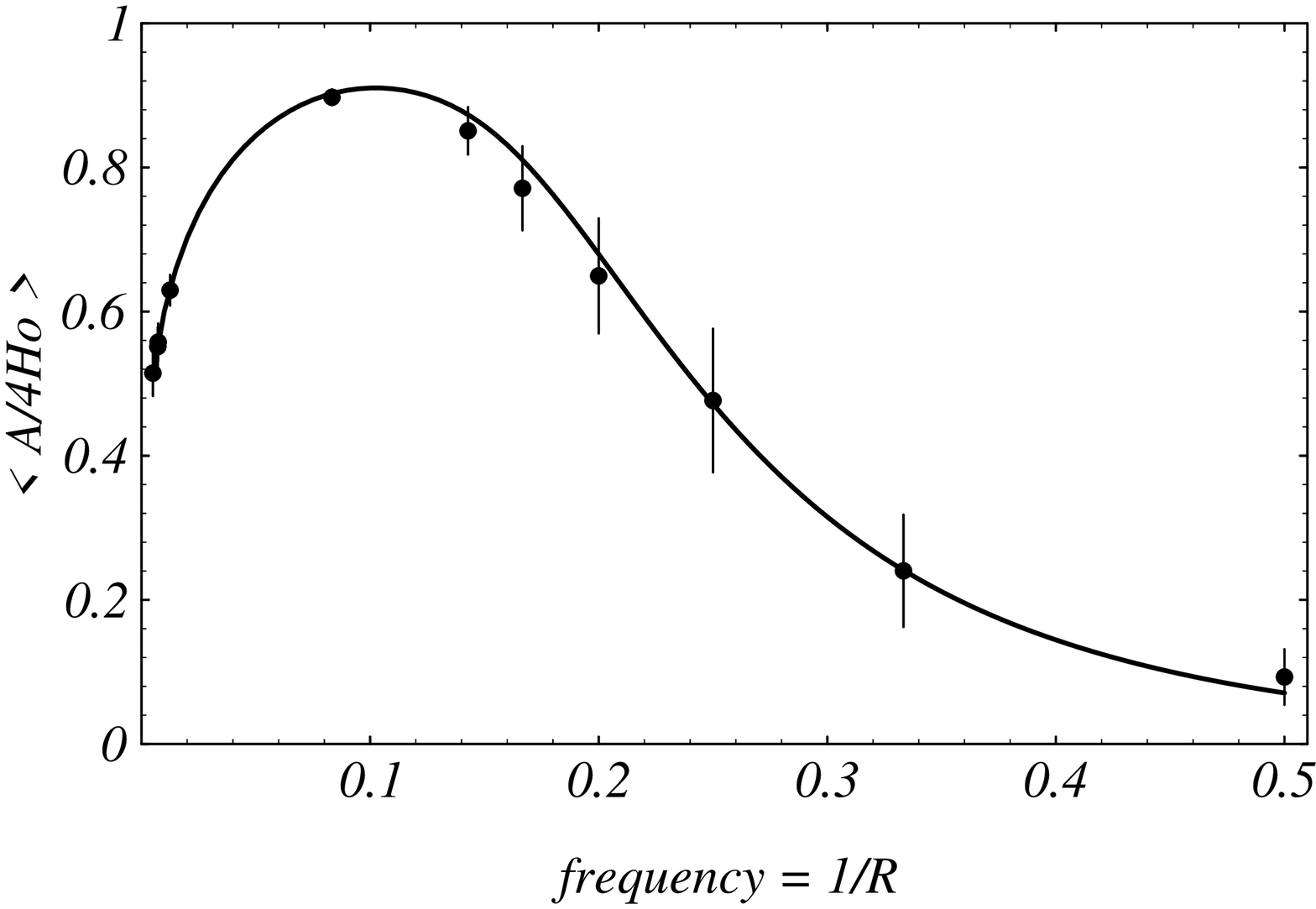,width=3.75in}}
\end{figure}
\pagebreak

\begin{figure}
\noindent
Figure 3
\hspace{5.0in}
{\large {\bf HD-11}} \\
S.W. Sides \\
MMM 96 \\
\centerline{\psfig{figure=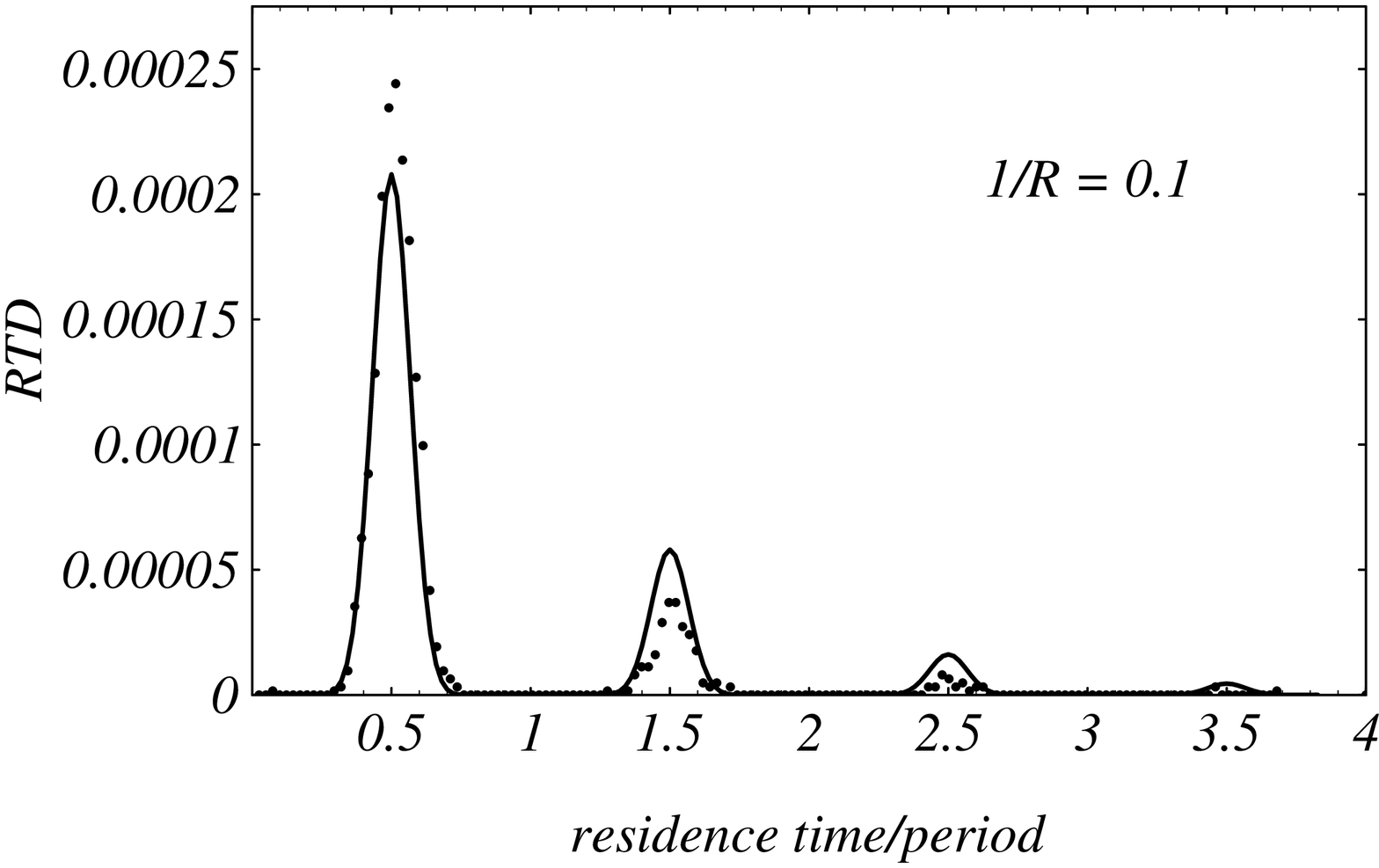,width=3.75in}}
\end{figure}
\pagebreak

\begin{figure}
\noindent
Figure 4
\hspace{5.0in}
{\large {\bf HD-11}} \\
S.W. Sides \\
MMM 96 \\
\centerline{\psfig{figure=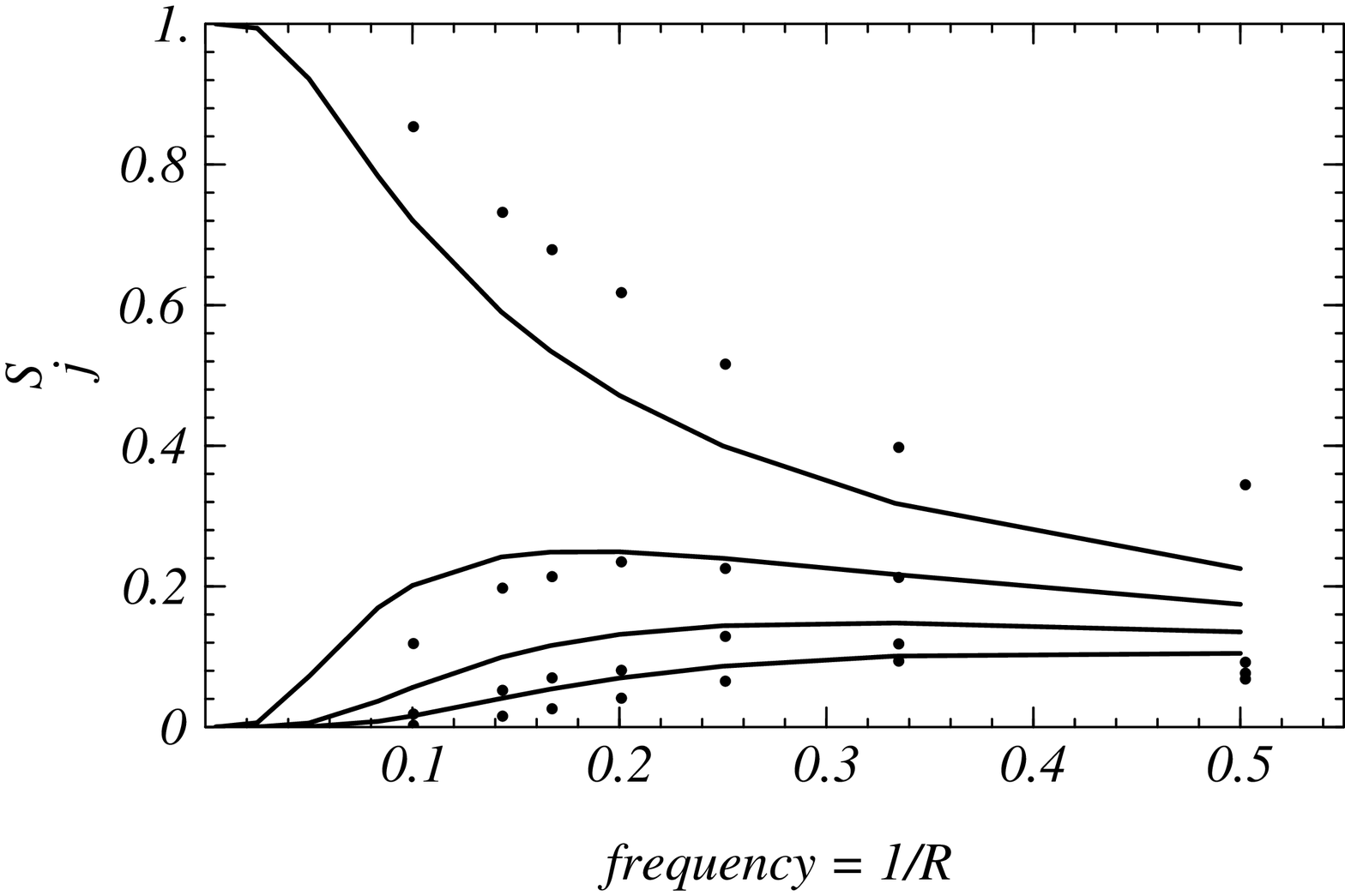,width=3.75in}}
\end{figure}
\pagebreak
%%%%%%%%%%%%%%%%%%%%%%%%%%%%%%%%%%%%%%%%%%%%%%%%%%%%%%%%%%%%%%%%%%%%%

%%\input{figs1}
%%\input{figs2}

\end{document}